\documentclass[aps,preprint]{revtex4-1}
\usepackage{graphicx}
\usepackage{amsmath}
\usepackage{amsfonts}
\usepackage{amsthm}
\usepackage{amssymb}
\usepackage{amsbsy}
\usepackage{wasysym}
\usepackage{bm}
\usepackage{mathrsfs}

\begin{document}

\title{Shear Induced Fluidization Of Thermal Amorphous Solids}

\author{Santhosh Kumar R}
\author {Bhaskar Sen Gupta}
\email{bhaskar.sengupta@vit.ac.in}
\affiliation{Department of Physics, School of Advanced Sciences, Vellore Institute of Technology, Vellore, Tamil Nadu - 632014, India}

\begin{abstract}
We study the shear-induced fluidization of amorphous solids subjected to external loading by investigating the relaxation dynamics of the deformed states using computer simulation. A simple shear deformation is employed at constant rate to the thermal glassy materials. The shear localization and the plastic deformation heterogeneity with strain is investigated in terms of the non-affine displacement field. The mean square displacement shows an enhanced mobility of the particles with strain, indicating the fluidization of the material. Using the time correlation function we estimate the relaxation time of the sheared glasses. A significant decrease in the relaxation time is observed up to the yielding point as the material loses its solid nature and eventually becomes liquid-like. Finally, the imprint of memory of the quiescent sample on the rheological properties of the shear melted glass is investigated by computing the relaxation of the shear stress. We find a finite persistent residual stress that outlasts the experimental observation time in our system. 
\end{abstract}

\maketitle


\section{Introduction}\label{sec1}
When a glass-forming liquid is cooled faster than its molecular relaxation time below the glass transition temperature $T_g$, amorphous solid is formed. This is essentially a frozen-in liquid state which exhibits a liquid-like organization of the constituent particles. But unlike liquids that flow under any amount of external strain, amorphous materials respond elastically to external small strain. As a result, the stress increases linearly with strain. Of course, this linear regime is punctuated with small localized plastic events \cite{06ML,Tanguy,Koumakis}. On further increase of strain, the system exhibits the so-called yielding transition \cite{Alexander, Bonn}. After that the material starts to flow plastically either homogeneously or by shear localization depending on the temperature, the shear rate, and the density of the material \cite{04VBB, 04ML, 05DA, 06TLB, 06ML, 09LP, 11RTV, 06SLG, 13KTG, 13NSSMM}. In this regime, the material behaves like a liquid and the stress remains constant on an average with strain. This phenomenon is very generic to the so-called soft materials including metallic glass, foam, gel, clay, and soil, observed in numerous experiments \cite{Divoux,Chikkadi,Besseling,Bokeloh} and computer simulations \cite{Varnik,Chaudhuri,Shi,Bailey,Tsamados,Tsamados1,Fusco}. 

The current state-of-the-art of these materials is however poorly understood \cite{sugg1,sugg2,sugg3}. Unlike the equations for fluid mechanics (e.g. Navier-Stokes equation) which describes the fluid flow, a successful theory for the steady-state elastoplastic flow is still lacking. This can be attributed to the large variety of micro-structures explored by these materials. Given the fact that glassy materials are out of equilibrium solids and at very low temperatures (below $T_g$), they are stuck in one of the metastable minima in the potential energy landscape where the constituent particles vibrate around some mean position. The configurations of a particular glassy sample will reside in close proximity over a long period (large compared to the experimental time scale). As a result, the system is non-ergodic and has a limited number of accessible microstates. The application of external loading distorts the energy landscape \cite{06ML}. It was shown in the athermal quasistatic limit that the yielding process unleashes a vast number of configuration states that were inaccessible before yielding \cite{Prabhat,murari,bhaskar1, bhaskar2}. This is often termed as strain-induced ergodization and the material fluidize under the influence of large strain. 

However, the above scenario is much more complex at finite temperature because of the interplay of thermal energy and external forces. External strain modifies the potential energy landscape which is accompanied by significant modification of the structure of the glassy material via non-affine displacements of particles as indicated by experimental observations \cite{Rosner,Schmidt}. These structural modifications in turn affect the characteristics of the material such as relaxation dynamics and mechanical properties. Therefore, understanding and characterizing the rheological response of amorphous materials is of fundamental interest in the field of science and engineering applications. In this context, the most important process that serves as a sensitive prove of interest is usually the degree of plastic localization and the relaxation dynamics of the sheared glass \cite{gaurav}. 

In this paper, we seek to understand the shear induced fludization of a generic modeled thermal amorphous solid under simple shear deformation at constant shear rate using computer simulation. For that, the non-affine displacements of particles are analyzed to understand the the microscopic details of the plastic events in the pre- and post-yield deformed states. The particles with large non-affine displacement are found to form clusters which are homogeneously distributed in space in the pre-yield regime. The mobility of these particles increases with strain. After the yielding transition the group of particles localize within a system-spanning shear band with enhanced mobility, underlining the material fluidization. The same is reflected in the probability distribution of the non-affine motion. To gain more insight, the spatio-temporal evolution of the system is explored by computing the mean square displacement (MDS) of the constituent particle. The shear-induced fluidization is demonstrated by studying the relaxation dynamics of our system at various deformed states in terms of the intermediate scattering function. Finally, the role of the preparation history of these materials on their mechanical properties is investigated by examining the stress-relaxation and the residual stress by allowing the deformed states to evolve in time. 

Given this background, the paper is structured as follows. In section 2, we outline the model and the numerical method to prepare the thermal glass, the deformation protocol at a constant shear rate, and the relaxation procedure of the sample at a fixed deformation. In section 3 we discuss the numerical results obtained from our simulation by computing various macroscopic quantities discussed above. Finally, in section 4 we offer a summary and a discussion of the results presented in this paper.

\section{Numerical Simulation Details}
To prepare the thermal glass we use the well studied Kob-Anderson binary mixture model \cite{Kob}. Our model consists of $N$ classical point particles confined in a three-dimensional simulation box in the $NVT$ ensemble.  We choose a binary Lennard-Jones (LJ) mixture of particles which are labeled as A and B, and their number ratio is $80:20$. For simplicity, the mass of both types of particles $m$ is taken to be the same and equal to unity. The interaction potential for a pair of particles has the following form,
\begin{eqnarray}
U_{\alpha\beta}(r) &=& 4\epsilon_{\alpha\beta}\Big[\Big(\frac{\sigma_{\alpha\beta}}{r}\Big)^{12} - \Big(\frac{\sigma_{\alpha\beta}}{r}\Big)^{6} + A_0 \nonumber\\ &+& A_1\Big(\frac{r}{\sigma_{\alpha\beta}}\Big) + A_2\Big(\frac{r}{\sigma_{\alpha\beta}}\Big)^2\Big] \ , r\le r_{cut} \nonumber \\
&=& 0 \,~~~~~~~~~~~~~~~~~~~~~~~~~~~~~~~~ {r\textgreater r_{cut}}, \label{Uij}
\end{eqnarray}
where $\alpha, \beta \in  \rm{A, B}$. The units of various quantities in our simulation are as follows: lengths are expressed in the unit of $\sigma_{AB}$, energies in the unit of $\epsilon_{AB}$, time in the unit of $(m\sigma_{AB}^2/\epsilon_{AB})^{1/2}$ and temperature in the unit of $\epsilon_{AB}/k_{\rm B}$. Here, $k_{\rm B}$ is the Boltzmann constant which is unity. The parameters $\sigma_{\alpha\beta}$ and $\epsilon_{\alpha\beta}$ are chosen as follows: $\sigma_{AA} = 1.0, \sigma_{BB} = 0.88, \sigma_{AB} = 0.8$ and $\epsilon_{AA} = 1.0, \epsilon_{BB} = 0.5, \epsilon_{AB} = 1.5$. With these set of parameters, a binary LJ mixture avoids crystallization below $T_g$ and forms stable glass. $T_g$ for this model is approximately $0.44$ in the reduced unit \cite{Kob}. To improve the computational efficiency, the potential is truncated at $r_{cut}=2.5\sigma_{AA}$. The parameters $A_0 = 0.040490$, $A_1 = -0.009702$ and $A_2 = 0.000620$ are such that the potential and its derivatives (up to second order) go to zero smoothly at $r=r_{cut}$. 

To prepare the glass samples, we use MD simulation. All simulations are carried out at density $\rho=1.2$ and a total number of particles $N=20000$. Position and velocity of particles are updated using the velocity-Verlet integration technique \cite{Verlet} with time step $\Delta t =0.005$. The temperature of the system is kept fixed to the desired value by employing the Berendsen thermostat \cite{Berendsen}. Also, we apply periodic boundary conditions in all directions. 

We begin our simulation by equilibrating the binary mixture in the liquid state at temperature $T=1.0$. To prepare the glassy sample, the system is then cooled to the final temperature $T=0.3$ below the glass transition temperature $T_g=0.43$ with a cooling rate $10^{-5}$. For statistical averaging, the whole process is repeated $100$ times with different initial realizations. Finally, the glassy sample is sheared along the $xz$ plane at a constant shear rate $\dot{\gamma}=10^{-4}$ up to $100 \%$ strain. During the shear process, the temperature is controlled by dissipative particle dynamics thermostat \cite{DPD}. The deformed configurations are saved at chosen strain values $\dot{\gamma}t$. The relaxation dynamics of these deformed states are studied by allowing the system to evolve with time (keeping $\dot{\gamma}t$) using MD simulation. The time $t$ is reset to zero at the beginning of the relaxation process.
\section{Results} 
In this section, we present the results obtained from our simulation. In Fig. 1 we show the typical stress vs strain curve when the external loading is imposed on the freshly quenched glass at a constant deformation rate. 
\begin{figure}[!ht]
	\centering
	\includegraphics[width=\columnwidth]{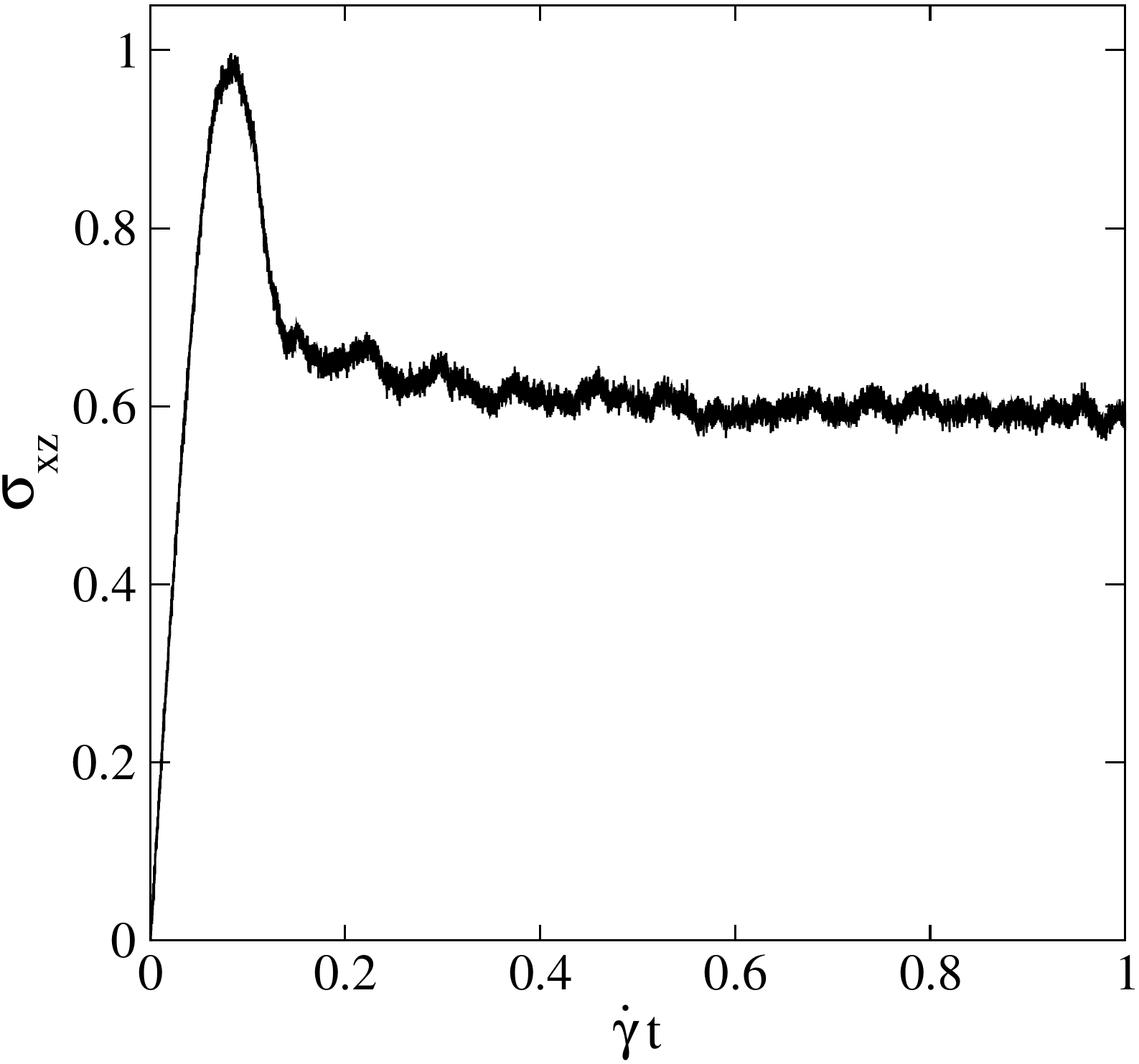}
	\caption{Typical stress $(\sigma)$ vs strain $(\dot{\gamma} t)$ curve of the glassy material of $N=20000$ particles at $T=0.3$ under simple shear deformation with constant shear rate $\dot \gamma=10^{-4}$ averaged over $100$ samples.}
	\label{fig1}
\end{figure}
As it is clearly observed, for the small shear deformation the stress increases linearly with strain and the system behaves as an elastic solid. Subsequently, the material yields with a stress overshoot. This is the hallmark of the loading behavior of almost all the soft materials. Of course, the yield stress depends on the history of the preparation of the samples. Upon further increase of strain, the steady-state plastic flow sets in, and the material behaves like a flowing fluid.  
\subsection{Plastic Deformation Heterogeneity}
As mentioned above, at finite temperature the shear-driven system undergoes structural changes due to non-affine rearrangements of cluster of particles with increasing strain. These irreversible structural rearrangements are assisted by the imposed shear deformation and thermal noise. In this section, we attempt to understand this heterogeneous flow of the system by analyzing the spatial organization of these clusters. This is best realized by the microscopic analysis involving spatial configurations of atoms with large relative non-affine displacement. Therefore, we compute the non-affine displacement of the local shear transformations in deformed solids defined as \cite{falk}
\begin{equation}
\begin{split}
D^2(t,\Delta t)  = & \frac{1}{N_i} \sum\limits_{j=1}^{N_i}  \lbrace \boldsymbol{\mathrm{r}}_j(t+\Delta t) - \boldsymbol{\mathrm{r}}_i(t+\Delta t) - \\
&   \boldsymbol{\mathrm{J}}_i [\boldsymbol{\mathrm{r}}_j(t) - \boldsymbol{\mathrm{r}}_i(t)]\rbrace^2
\end{split}
\label{eq-d2}
\end{equation}
Here $\boldsymbol{\mathrm{r}}_i(t)$ is the position vector of the $i$th particle at time $t$, $\boldsymbol{\mathrm{J}}_i$ is the transformation matrix that maps the $i$th particle and its neighbor (within the cutoff $r_{cut}$) at time $t$ and $t+\Delta t$ respectively via an affine deformation. This quantity is extensively used for the spatio-temporal analysis of non-affine displacements in shear-driven amorphous solids \cite{jana,chikkadi,varnik,ding}. 

\begin{figure*}[h!]
	\centering
	\begin{tabular}{cc}
		\centering
		\includegraphics[width=70mm]{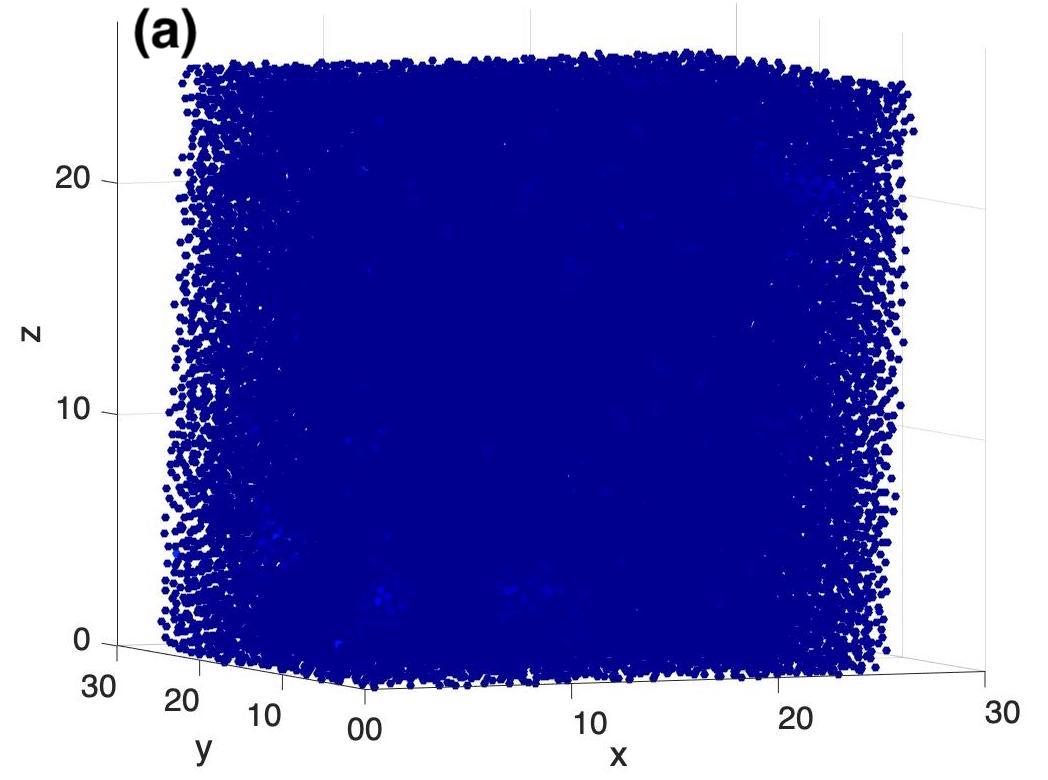}&
		\includegraphics[width=70mm]{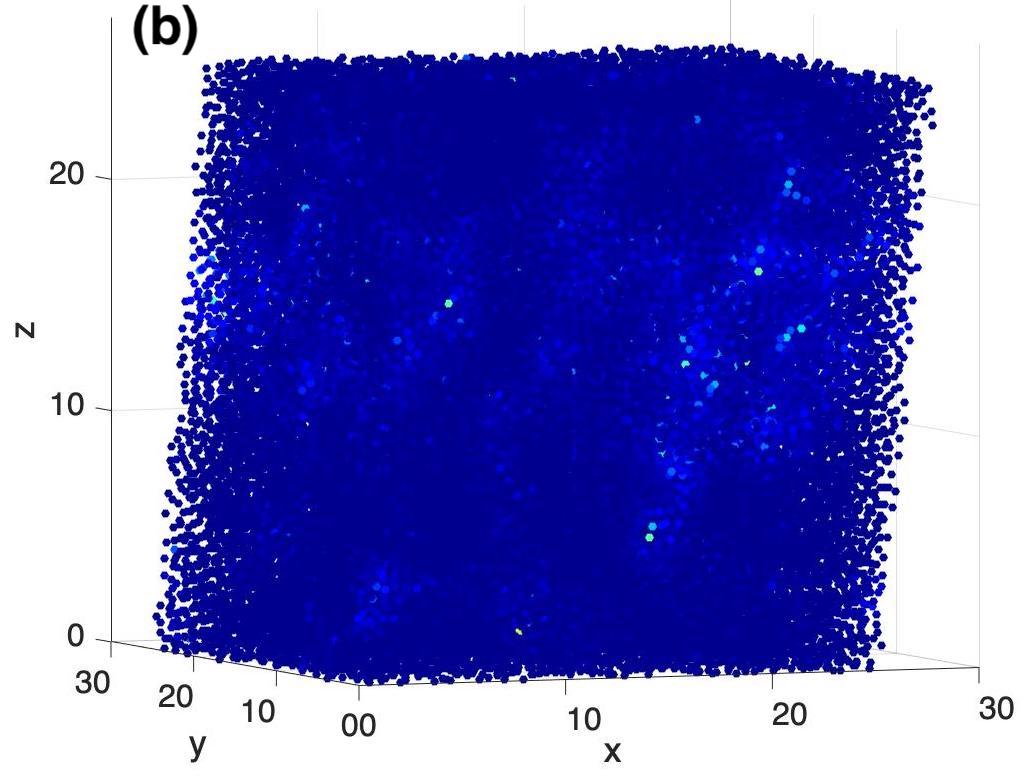}\\
		\includegraphics[width=70mm]{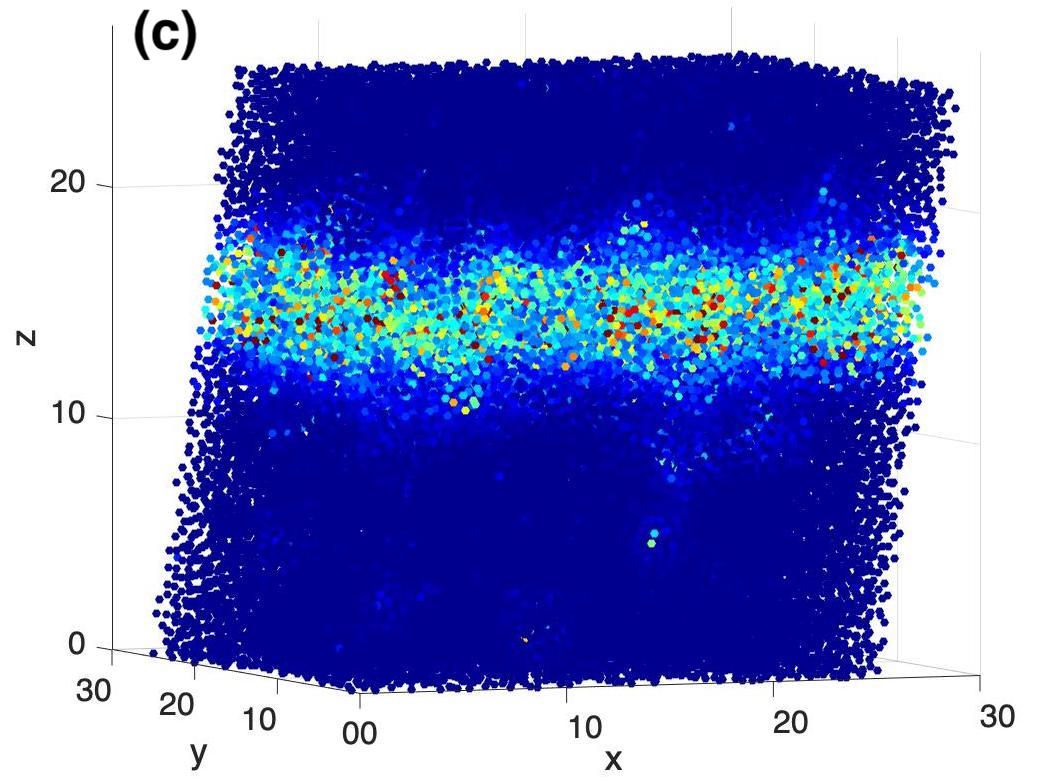}&
		\includegraphics[width=70mm]{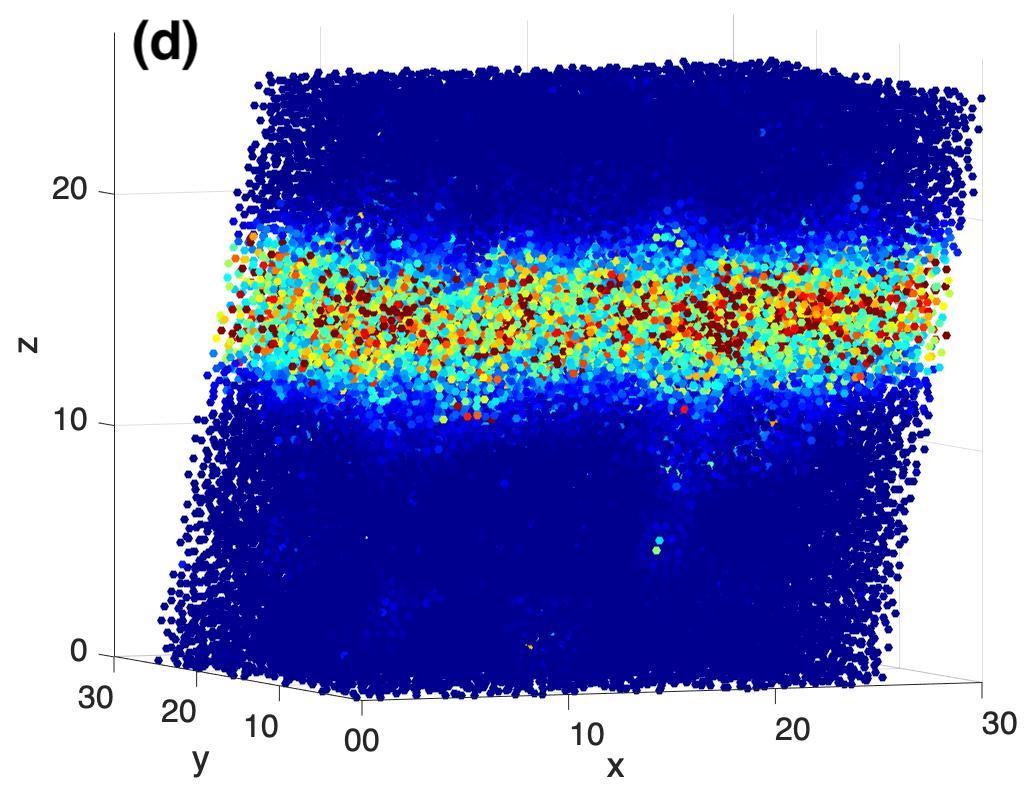}\\
		\includegraphics[width=68mm]{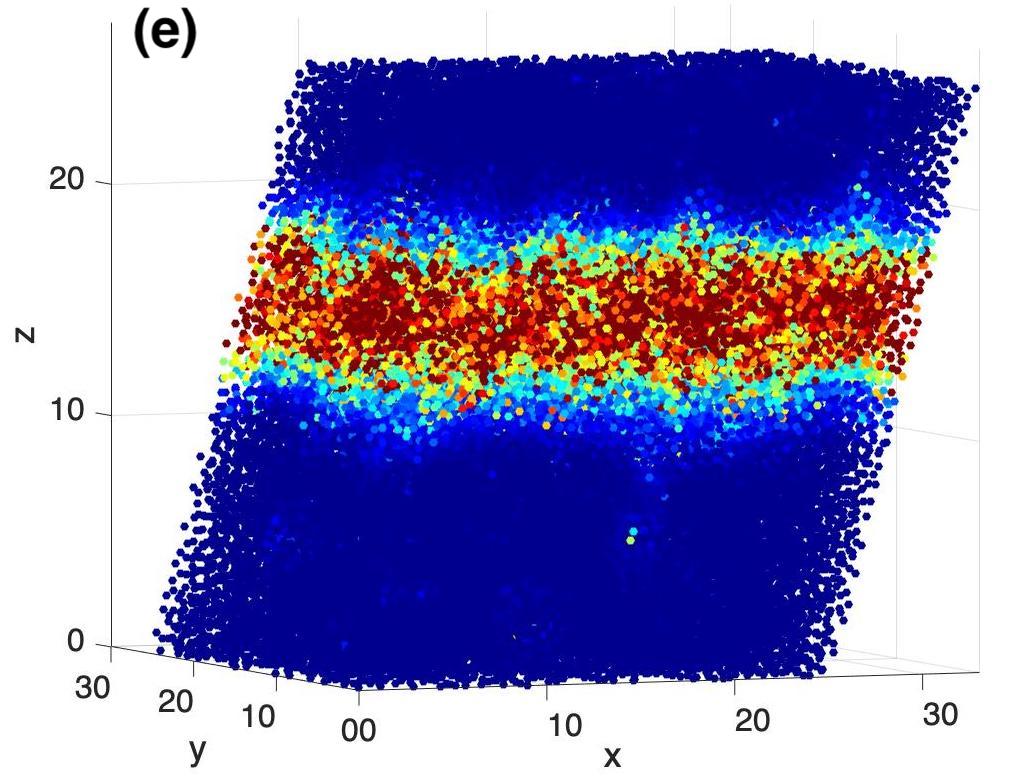}&
		\includegraphics[width=75mm]{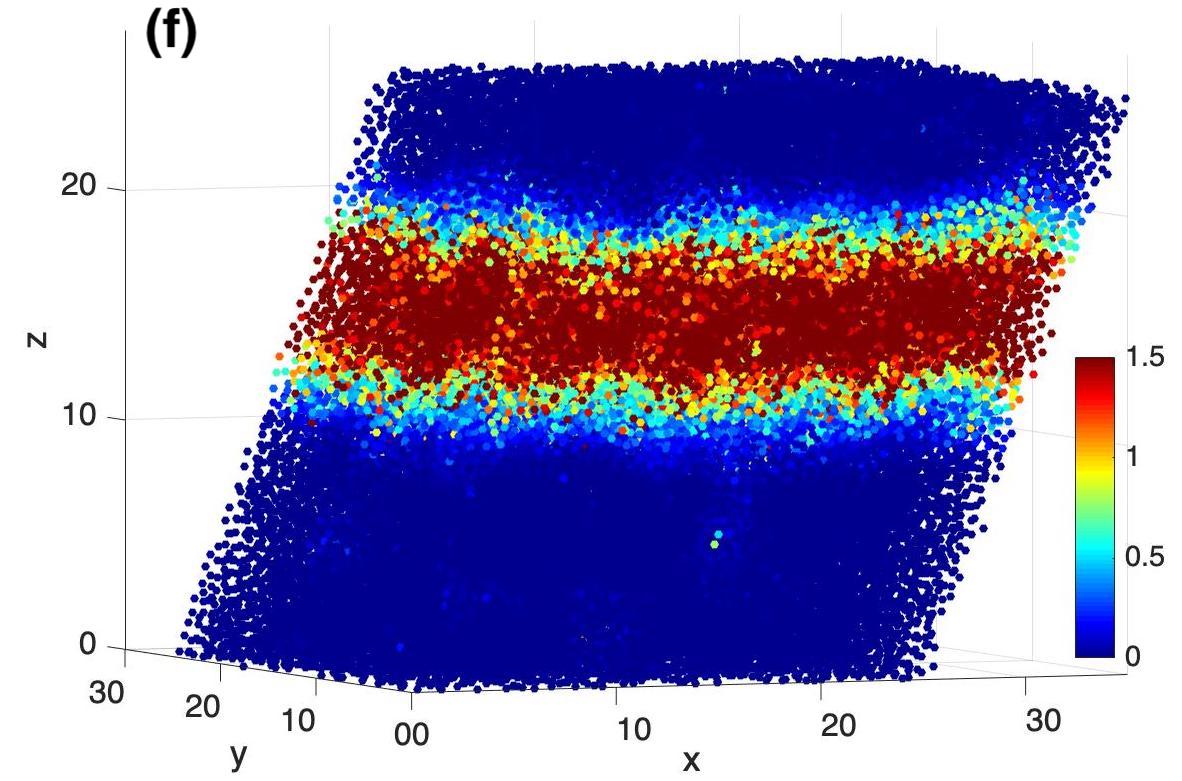}\\
	\end{tabular}
	\caption{Typical snapshots of the deformed glassy samples at the strain values (a) 0.05, (b) 0.10, (c) 0.15, (d) 0.20, (e) 0.30 and (f) 0.40. The particles are color-coded according to their non-affine displacements $D^2(t,\Delta t)$ (see text). The color coding scale is given on the right of figure (f).}
	\label{d2-snap}
\end{figure*}

\begin{figure}[h!]
	\centering
	\includegraphics[width=\columnwidth]{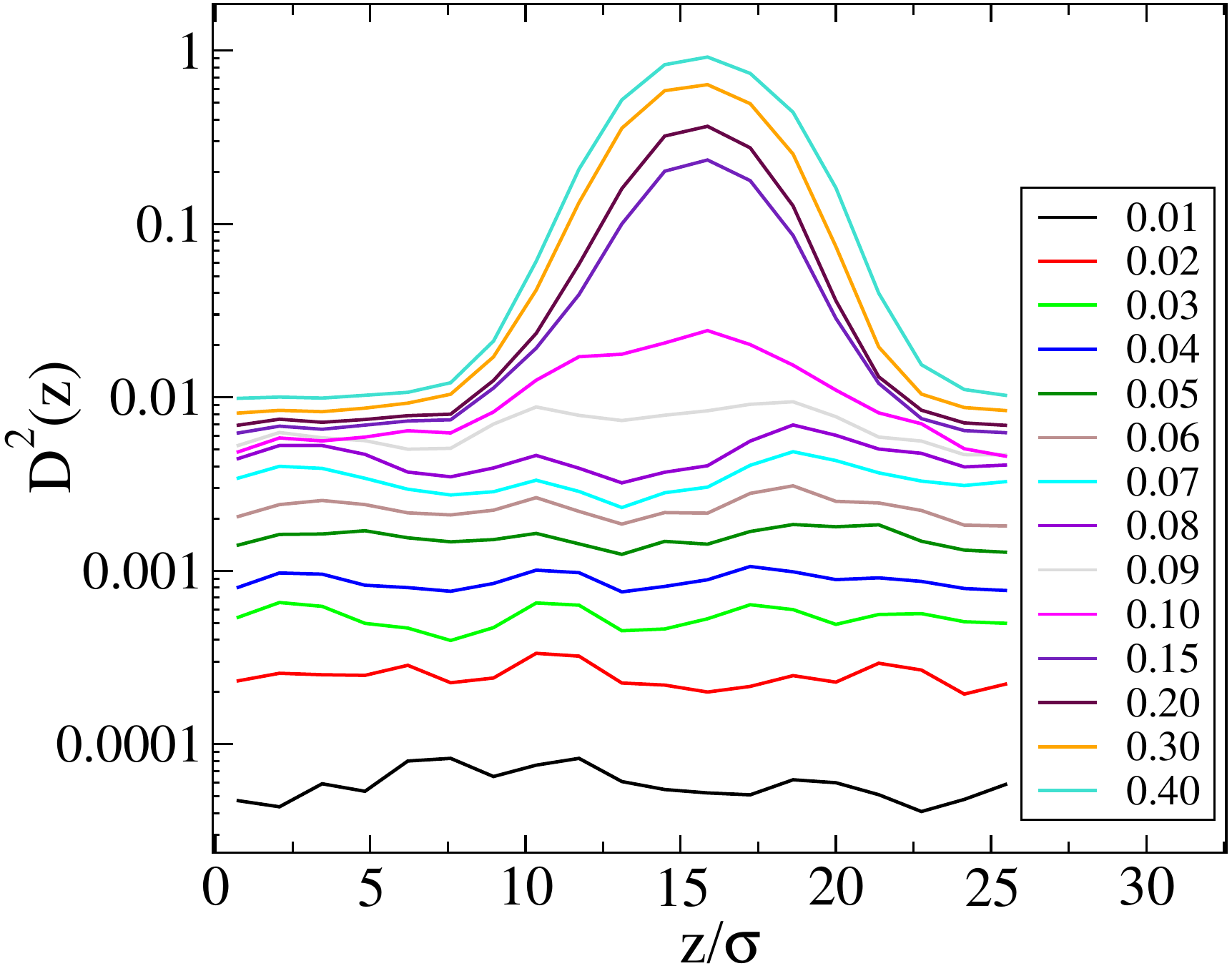}
	\caption{The spatial distribution of the non-affine displacement $D^2(z)$ vs $z$ at different strain values. The $D^2(z)$ shown here corresponds to the same sample shown in Fig. \ref{d2-snap}.}
	\label{d2_dist}
\end{figure}

In Fig. \ref{d2-snap} we show the snapshots of the deformed glassy samples for the shear strain $\dot \gamma t=$ 0.05, 0.1, 0.2, and 0.3. We compute the $D^2(t,\Delta t)$ from Eq. \ref{eq-d2} for each of these cases with respect to the undeformed atomic configuration. In Fig. \ref{d2-snap} the particles are color-coded according to their non-affine displacement field $D^2(t,\Delta t)$. From the figure, it is conspicuous that in the pre-yield regime ($\gamma=0.05$) the shear deformation is pretty homogeneous. As the system undergoes yielding transition, the mobile particles localize in a band-type structure running across the system and form shear band. The typical width of the shear band is of several particle diameters. Careful observation reveals that this width increases with increase in strain. This can be better understood by the probability distribution of the displacement field which involves spatially averaged profiles of $D^2(z)$ computed over a small bin of thickness $\Delta z$. This is shown in Fig. \ref{d2_dist} for a chosen set of strain values. For small deformation, the distribution curves are almost flat, indicating the cluster of mobile particles are homogeneously distributed across the sample. The average value of $D^2(z)$ increases gradually with strain. The yielding transition is marked by the appearance of a distinct peak in the distribution, indicating the formation of shear band. The width of the peak increases gradually with strain implying the expansion of the shear band. These results clearly indicate the enhanced mobility of the particles within the band and the material fluidization.

\subsection{The Mean Square Displacement}
Now we focus on the response of the material at various external loading conditions. For that, the deformed configurations saved at specific strain values are allowed to relax in time using the MD simulation. To obtain the information about the mobility of particles in the deformed states we compute the mean-square displacement (MSD) in the direction transverse to the applied shear defined as
\begin{equation}
\langle \Delta r_z^2(t) \rangle = \frac{1}{N_A} \sum\limits_{j=1}^{N_A}  \langle|r_{z,j}(t)-r_{z,j}(0)|^2\rangle
\end{equation}
where $N_A$ is the number of $A$-type particles and $r_{z,j}(t)$ is the z-component of the position vector of the particle $j$ at time $t$. This is shown in Fig. 2 for various deformed states. 
\begin{figure}[!ht]
	\centering
	\includegraphics[width=\columnwidth]{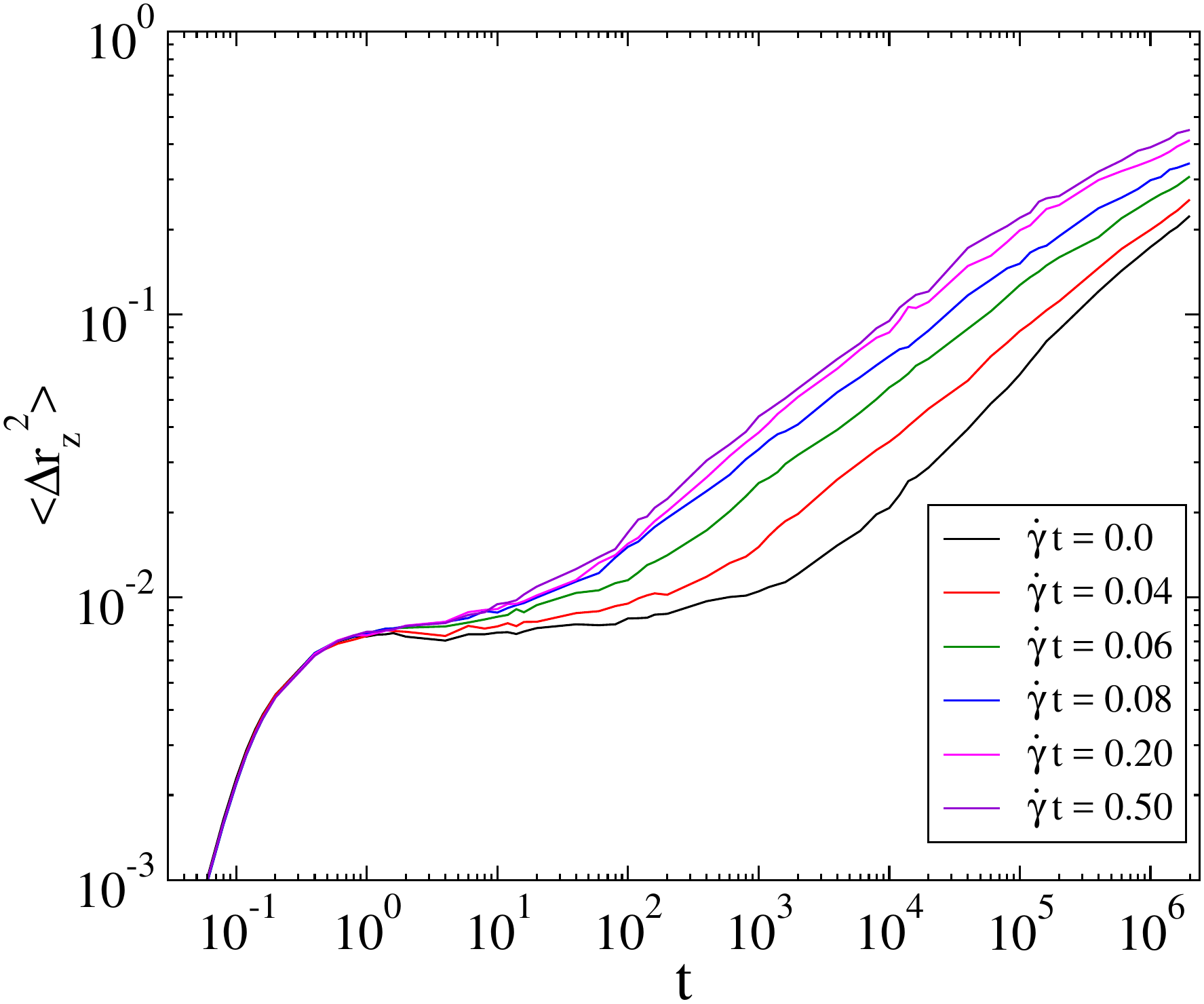}
	\caption{The evolution of the averaged mean square displacement $\langle \Delta r_z^2(t) \rangle$ with time for the deformed systems corresponding to the strain values $\dot \gamma t$ = 0.0, 0.04, 0.06, 0.08, 0.2 and 0.5 (see text).}
	\label{fig2}
\end{figure}
The MSD gives the typical average distance that a particle moves within a time $t$. For short times we see the ballistic behavior with $\langle r(t^2) \rangle \sim t^2$ \cite{Hansen}. In the undeformed state ($\dot{\gamma} t = 0\% $), unlike liquid, the MSD does not enter into the diffusive regime but instead shows a plateau at intermediate time. This is attributed to the so-called "cage effect" where the particle temporarily gets trapped by its neighbor. After a sufficiently long time, the cage is broken by thermal energy and the MSD crosses over to a time-dependent diffusive regime. 

However, the situation changes for the finite deformation cases. With increasing strain, the plateau region becomes narrower and the cage effect becomes weaker. This is an indication of the material losing its solid-like behavior. We have checked that the existence of plateau is observed up to the yielding transition point. In the post-yield relaxation, the plateau is short-lived and the MSD almost directly crosses over from ballistic to the diffusive regime. The enhanced mobility in this regime is attributed to the large plastic drops via the system spanning shear band formation inside the material and the non-affine displacement of a large number of particles. Our results on MSD are in agreement with the observations in recent experimental and numerical studies \cite{gaurav}. At very high deformation the MSD curves almost overlap with each other. These results indicate the fluidization of the material under the effect of shear.
\subsection{Time Correlation Function}
\begin{figure}[!ht]
	\centering
	\includegraphics[width=\columnwidth]{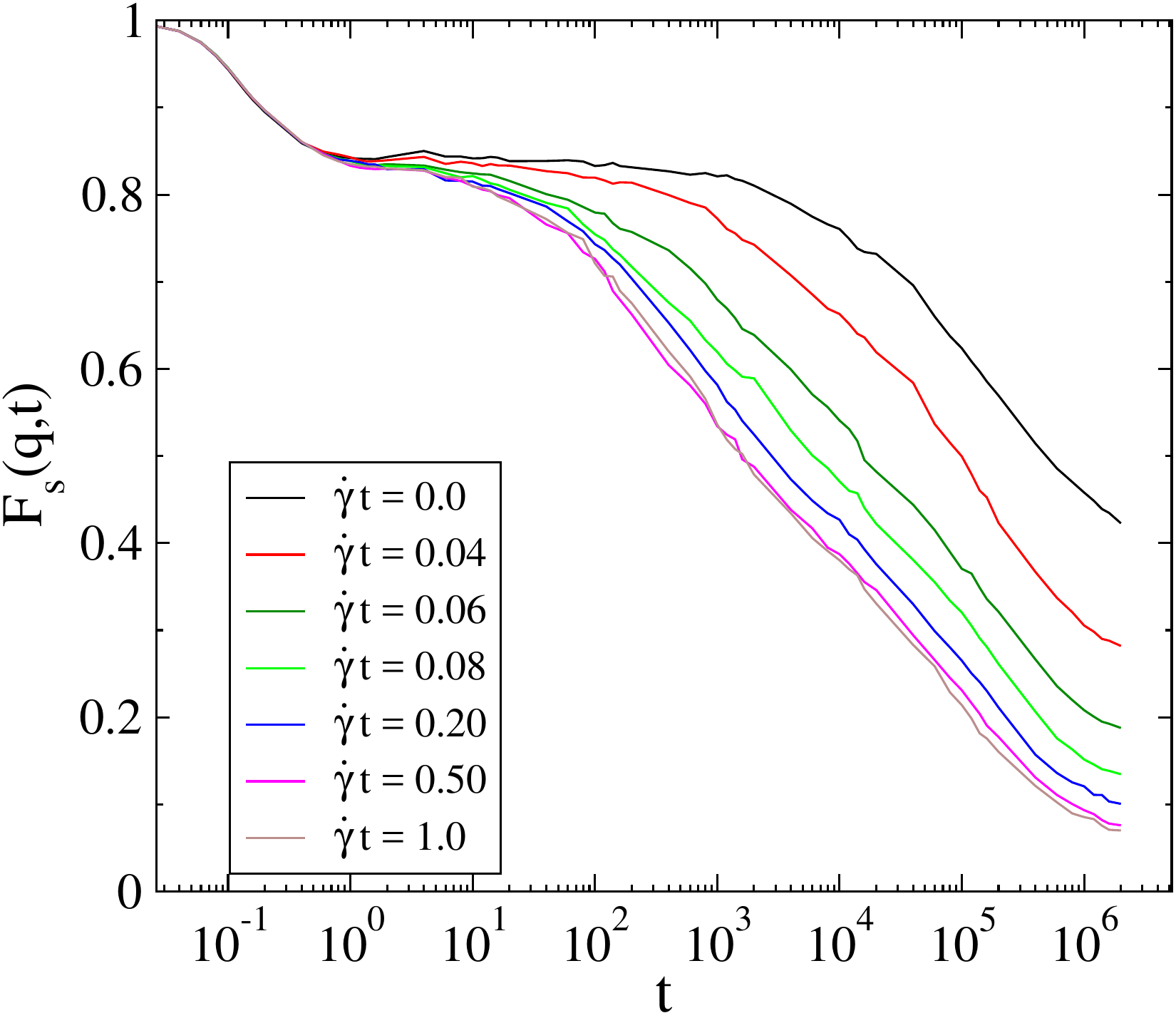}
	\caption{The time evolution of the correlation function $F_s(q,t)$ for the deformed systems corresponding to the strain values $\dot \gamma t$ = 0.0, 0.04, 0.06, 0.08, 0.2, 0.5 and 1.0.}
	\label{fig3}
\end{figure}

To gain more insight into the relaxation dynamics we focus on the typical time correlation function. Therefore, we compute the self part of the intermediate scattering function defined as
\begin{equation}
F_s(q,t) = \frac{1}{N_A} \left\langle \sum\limits_{i=1}^{N_A} e^{i.q\lbrack\vec{r}_j(t)-\vec{r}_j(0)\rbrack}\right\rangle
\end{equation}
where $q=7.25$ is the wave vector at the structure factor peak of $A$ type of particles \cite{kob1}.  The $F_s(q,t)$ is shown in Fig. \ref{fig3} for the same set of strain values as in Fig. \ref{fig2}. 
For the undeformed glassy sample, we observe the typical two-step relaxation. Here also we find the ballistic regime for the short-time dynamics. This is followed by a plateau in the correlation function. The time window over which the plateau is observed is called the $\beta$-relaxation \cite{gotze}. This physical meaning of the $\beta$-relaxation is given by the so-called cage effect, explained in the previous section. Finally, at long time (logarithmic time scale) the correlation decays to zero via the $\alpha$-relaxation. In the deformed system, we observe that $F_s(q,t)$ decays faster with increase in strain. After the yielding transition, the $\beta$-relaxation part becomes short-lived. At sufficiently large deformation, the correlation function becomes almost insensitive to the strain value. This observation is consistent with the MSD behavior shown in Fig. \ref{fig2}

\begin{figure}[!ht]
	\centering
	\includegraphics[width=\columnwidth]{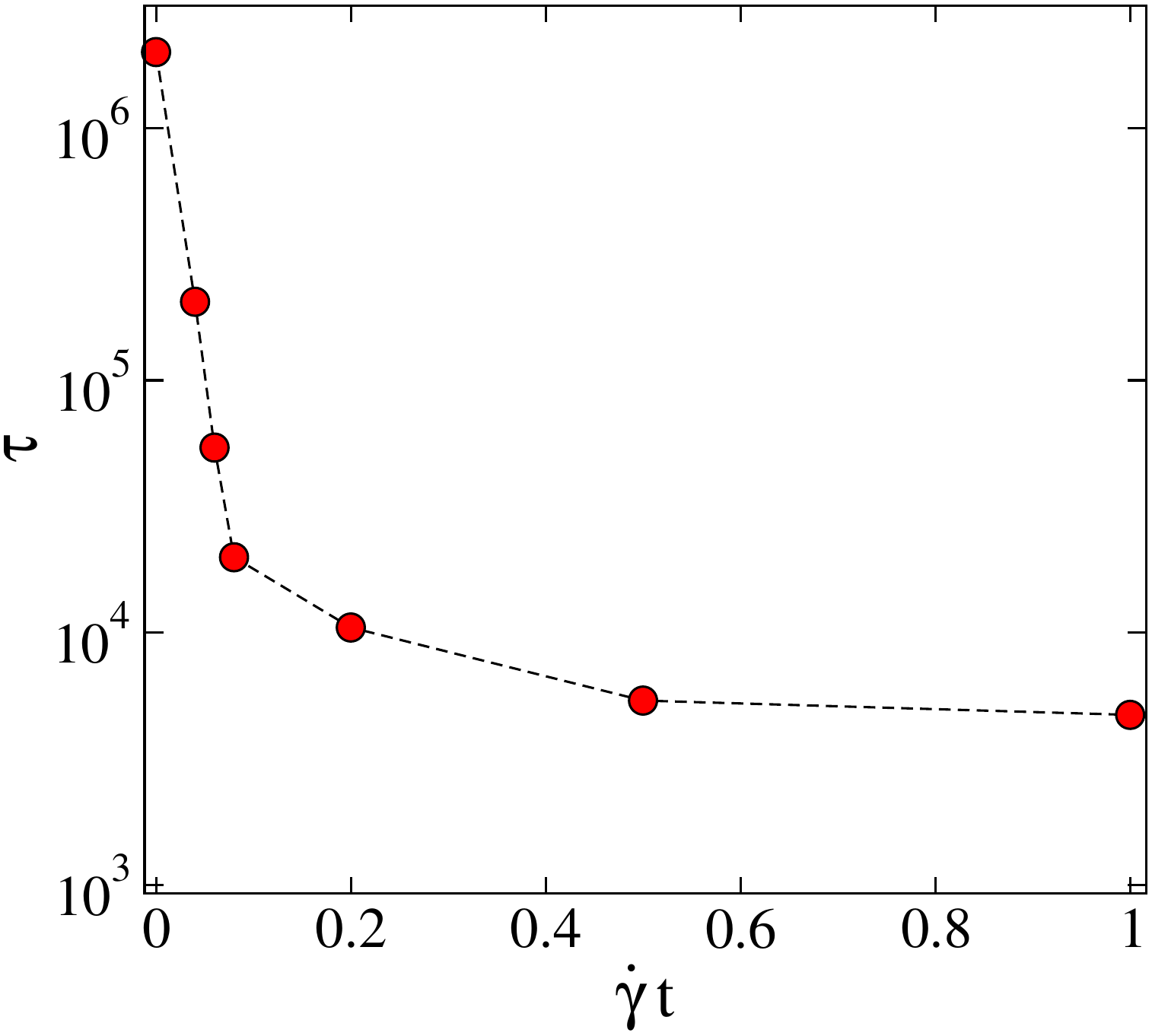}
	\caption{The relaxation time $\tau_r$ obtained from $F_s(q,t)$ as a function of $\dot \gamma t$.}
	\label{fig4}
\end{figure}

To quantify the decay dynamics of $F_s(q,t)$ with strain, we now compute the relaxation time $\tau_r$ associated with the correlation function. The $\tau_r$ is defined as the time it takes to reach the value $F_s(q,t)=0.42$ which is the lowest value of $F_s(q,t)$ for $\dot{\gamma} t=0$. The variation of $\tau_r$ with $\dot{\gamma} t$ is displayed in Fig. \ref{fig4}. For the quiescent sample, the relaxation is the slowest with maximum $\tau_r$. On increase of strain $\tau_r$ decreases rapidly up to the yielding point. After that, $\tau_r$ becomes almost constant in the steady-state plastic flow regime. The rapid decrease of the relaxation time in the linear regime of the loading curve and eventually become saturated at a lower value under the influence of large deformation is a clear indication of strain-induced fluidization of the material. 

It is worth noting that the characteristics of the correlation function displayed here at high strain do not exactly resemble the behavior of a liquid. For a liquid above melting temperature, the $F_s(q,t)$ shows an exponential decay after the ballistic regime and the $\beta$-relaxation part is completely absent \cite{Hansen}. In case of maximum deformation ($\dot{\gamma} t = 100\%$) used in our simulation, the $F_s(q,t)$ shows a signature of two-step relaxation where the $\beta$-relaxation is short-lived but present. Also, the $\tau_r$ is much higher than the normal liquid. Therefore, the shear molten state in the steady flow regime resembles the nature of viscous supercooled liquids. But unlike the thermally quenched supercooled liquids, these states can not be characterized by its thermodynamic control variables because of the memory effect of the quiescent sample that fades away with strain but is never demolished by any amount of deformation \cite{gaurav,ballauff} as demonstrated in the next section.

\subsection{The Residual Stress}
\begin{figure}[!ht]
	\centering
	\includegraphics[width=\columnwidth]{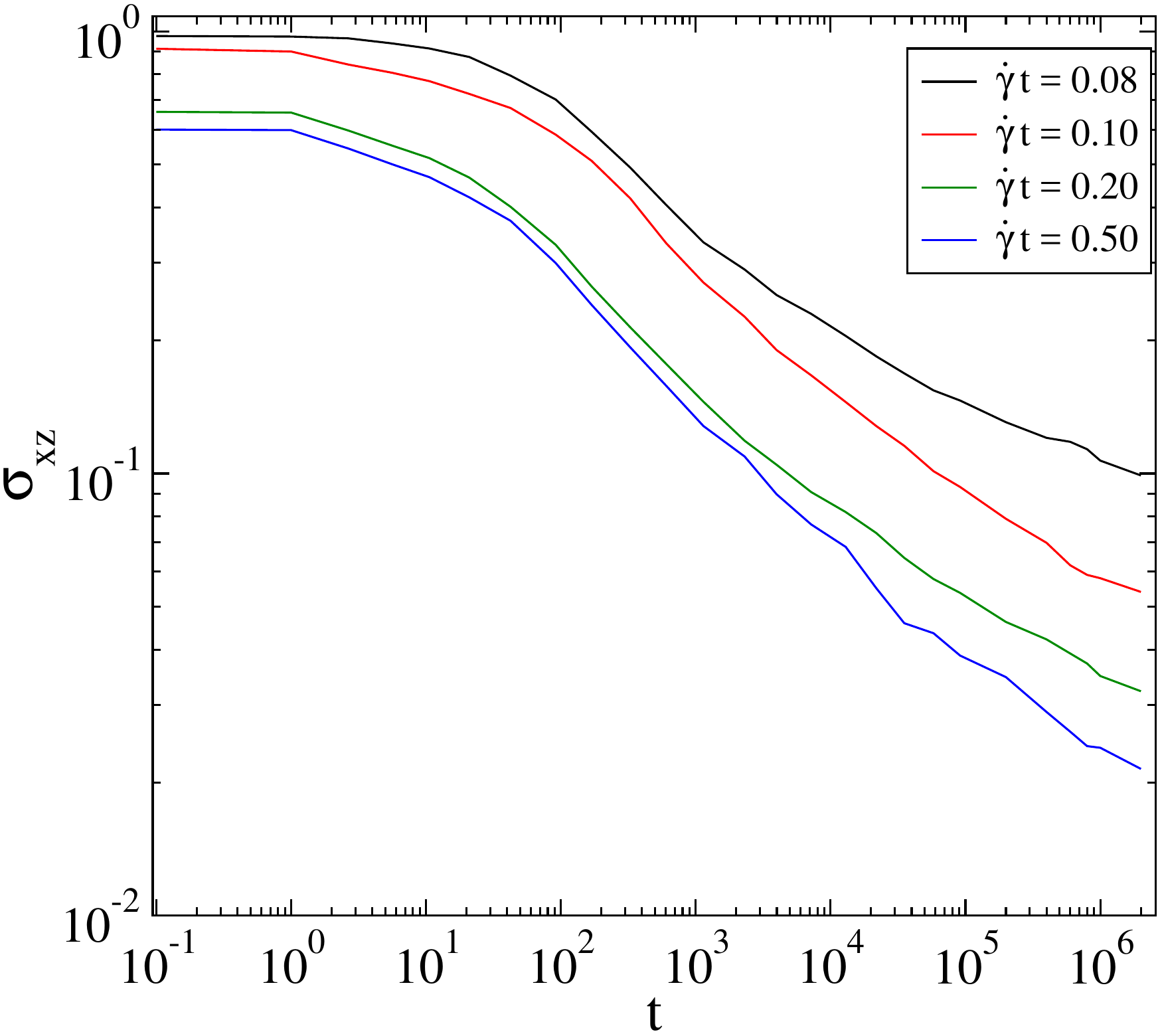}
	\caption{The relaxation dynamics of the stress $\sigma_{xz}$ with time $t$ for the deformed samples with $\dot \gamma t$ = 0.08, 0.1, 0.2 and 0.5.}
	\label{fig5}
\end{figure}

When the strain is imposed, the stress is build up in the glassy material as reflected in the loading curve. In this section, we investigate how this shear stress relaxes when the deformation is switched off at a chosen strain value and the system is allowed to evolve in time. The main purpose of this particular probe is to gain insight into the memory effect enshrined in the micro-structures of the deformed system due to the non-equilibrium thermal quench as a part of the sample preparation protocol. The stress relaxation is displayed in Fig. \ref{fig5} for a set of chosen strain values.  
At the beginning of the relaxation process ($t=0$) for a given strain, the stress is the same as reflected in the loading curve in Fig. \ref{fig1}. For short time, the stress remains almost constant for all the strain values and then starts to decrease. We also observe in Fig. 5 that the long time tail part of these curves shows a power-law decay (linear behavior in the log-log scale). Therefore, unlike a liquid where the stress becomes zero on the structural-relaxation time scale, the build-up stress in the deformed glass never relaxes to zero. In other words, the sheared configuration can never relax to equilibrium completely. As a result, we always find a persistent residual stress within the experimental time window. These results obtained from our simulation are very generic and observed in many soft materials like colloids, molecular glass, etc via experiments \cite{Dingenouts,Siebenburger,Koumakis,Koumakis1,Zausch} and simulations \cite{gaurav,ballauff}. The mode-coupling theory of glass transition also predicts the shear residual stress in glass dependent on shear history \cite{Gotze1,Fuchs}. The upshot of this section is that the memory inherited by the sample from the preparation protocol persists forever and affects the mechanical properties and relaxation dynamics in the sheared configuration. Therefore, the glassy sample in the plastic flow regime can not be completely characterized by the associated thermodynamics variables only. 
\section{Summary and Conclusions}
In this paper, we studied systematically the fluidization of amorphous material under the influence of external shear in terms of the relaxation dynamics under different loading conditions using computer simulation. Analyzing the non-affine displacement field it was shown that the local shear transformation zones are homogeneously distributed in space until the yielding transition and subsequently localizes in the form of system spanning shear band with enhanced mobility. The time evolution of the microstructure in both the elastic and plastic branches of the loading curve was studied. Using MSD we showed that the diffusivity of the constituent particles enhances with strain and the yielding transition plays a pivotal role in this. At high strain value, the system becomes fluid-like and the dynamics become diffusive at short time after the ballistic regime. The time scale of relaxation was estimated using the standard time correlation function. A dramatic decrease in the relaxation time is observed with strain up to the yielding point, after which it gradually saturates in the plastic region. These results clearly demonstrated that the solid (elastic) nature of the material was gradually lost under loading and became fluid-like after yielding. The role of the memory effect of the quiescent configuration in the fluidization process was also investigated via the shear stress relaxation with time. We observed that the stress did not decay to zero within the experimental time scale. Therefore, the rheological properties of the shear melted glass depend on the preparation history of the pre-sheared amorphous solids. It will be worth extending our work to other soft materials with more complicated nonlinear strain memories.

\section*{Acknowledgement} Bhaskar Sen Gupta acknowledges Science and Engineering Research Board (SERB), Department of Science and Technology (DST), Government of India (no. SRG/2019/001923) for financial support.

\end{document}